# Complementary Dual Subfield Linear Codes Over Finite Fields

**Kriangkrai Boonniyoma[†] and Somphong Jitman[‡,1]**

[†]Department of Mathematics, Faculty of Science, Silpakorn University,
Nakhon Pathom 73000, Thailand
e-mail : krai_b_555@hotmail.com

[‡]Department of Mathematics, Faculty of Science, Silpakorn University,
Nakhon Pathom 73000, Thailand
e-mail : jitman_s@silpakorn.edu

**Abstract :** Two families of complementary codes over finite fields $\mathbb{F}_q$ are studied, where $q = r^2$ is square: *i*) Hermitian complementary dual linear codes, and *ii*) trace Hermitian complementary dual subfield linear codes. Necessary and sufficient conditions for a linear code (resp., a subfield linear code) to be Hermitian complementary dual (resp., trace Hermitian complementary dual) are determined. Constructions of such codes are given together their parameters. Some illustrative examples are provided as well.

**Keywords :** Complementary dual codes; Hermitian inner product; trace Hermitian inner product; subfield linear codes.

**2010 Mathematics Subject Classification : 94B05, 12E20.**

## 1 Introduction

Linear codes with Euclidean Complementary dual have been studied in [3]. The characterization and properties of such codes were given. These codes are interesting since they reach the maximum decoding capability of adder channel [3]. Moreover, in some cases, such codes can be decoded faster than other linear codes using nearest neighbor decoding. In [6], necessary and sufficient conditions for cyclic codes to be Euclidean complementary dual have been determined. Hermitian complementary dual cyclic codes over finite fields have been characterized in [5].

---

[1]Corresponding author.

Subfield linear codes and their duals under the trace Hermitian inner product have been studied in [1] and [4]. Such codes have an application in constructing quantum codes in [1] and references therein.

To the best of our knowledge, Hermitian complementary dual linear codes and trace Hermitian complementary dual subfield linear codes have not been well studied. Therefore, it is of natural interest to studied complementary dual codes with respect to the Hermitian and trace Hermitian inner products. In this paper, we focus on Hermitian complementary dual linear codes and trace Hermitian complementary dual subfield linear codes. Characterizations, properties, and constructions of such codes are studied.

The paper is organized as follows: Some basic concepts and preliminary results on complementary dual codes are recalled in Section 2. In Section 3, characterization of complementary dual codes with respect to the two inner products are given. Some constructions and illustrative examples of such complementary dual codes are established in Section 4.

## 2 Preliminaries

Let $r$ and $q = r^2$ be prime power integers and let $\mathbb{F}_r \subseteq \mathbb{F}_q$ be finite fields. Let $\text{Tr} : \mathbb{F}_q \to \mathbb{F}_r$ denote the *trace map* given by $\text{Tr}(\beta) = \beta + \beta^r$ for all $\beta \in \mathbb{F}_q$. Some properties of the trace map can be found in [2, Theorem 2.23]. For $\mathbf{u} = (u_1, u_2, \ldots, u_n) \in \mathbb{F}_q^n$, let $\overline{\mathbf{u}} = (\overline{u_1}, \overline{u_2}, \ldots, \overline{u_n})$, where $\overline{a} = a^r$ for all $a \in \mathbb{F}_q$. For each matrix $A = [a_{ij}] \in M_{m,n}(\mathbb{F}_q)$, let $\overline{A} = [\overline{a_{ij}}]$ and $\text{Tr}(A) = [\text{Tr}(a_{ij})]$. Given $\mathbf{u}, \mathbf{v} \in \mathbb{F}_q^n$, let $wt(\mathbf{v})$ denote the *Hamming weight* of $\mathbf{v}$ and let $d(\mathbf{u}, \mathbf{v})$ denote the *Hamming distance* between $\mathbf{u}$ and $\mathbf{v}$.

A *code* of length $n$ over $\mathbb{F}_q$ is defined to be a nonempty subset $C$ of $\mathbb{F}_q^n$. The *minimum distance* $d(C)$ is given by

$$d(C) = \min\{d(\mathbf{u}, \mathbf{v}) \mid \mathbf{u}, \mathbf{v} \in C, \mathbf{u} \neq \mathbf{v}\}.$$

An $[n,k,d]_q$ *linear code* $C$ is a $k$-dimensional $\mathbb{F}_q$-subspace of $\mathbb{F}_q^n$ with minimum distance $d$. A $k \times n$ matrix $G$ over $\mathbb{F}_q$ is called a *generator matrix* for an $[n,k,d]_q$ linear code $C$ if the rows of $G$ form a basis of $C$.

For a general code $C \subseteq \mathbb{F}_q^n$, the notation $(n, M = |C|, d)_q$ is commonly used. A code $C$ is said to be an $\mathbb{F}_r$-*linear code* over $\mathbb{F}_q$ if $C$ is a subspace of the $\mathbb{F}_r$-vector space $\mathbb{F}_q^n$. When $r$ is clear from the context, $C$ is called a *subfield linear code* over $\mathbb{F}_q$. It is not difficult to see that if $C$ is an $\mathbb{F}_r$-linear code of length $n$ over $\mathbb{F}_q$, then $|C| = r^\ell$ for some $0 \le \ell \le 2n$ and $\dim_{\mathbb{F}_r}(C) = \ell$. An $\ell \times n$ matrix $G$ over $\mathbb{F}_q$ is called a *generator matrix* for an $(n, r^\ell, d)_q$ $\mathbb{F}_r$-linear code $C$ if the rows of $G$ form a basis of $C$ as an $\mathbb{F}_r$-vector space.

For $\mathbf{u} = (u_1, u_2, \ldots, u_n)$ and $\mathbf{v} = (v_1, v_2, \ldots, v_n)$ in $\mathbb{F}_q^n$, the inner products between $\mathbf{u}$ and $\mathbf{v}$ are defined as follows:

(1) $\langle \mathbf{u}, \mathbf{v} \rangle_{\mathrm{E}} := \sum_{i=1}^n u_i v_i$ is the *Euclidean inner product* of $\mathbf{u}$ and $\mathbf{v}$.

(2) $\langle \mathbf{u}, \mathbf{v} \rangle_{\mathrm{H}} := \sum_{i=1}^n u_i \overline{v_i} = \langle \mathbf{u}, \overline{\mathbf{v}} \rangle_{\mathrm{E}}$ is the *Hermitian inner product* of $\mathbf{u}$ and $\mathbf{v}$.

(3) The *trace Hermitian inner product* are defined into two cases depending on the field characteristic:

  (a) For an even $q$, $\langle \mathbf{u}, \mathbf{v} \rangle_{\mathrm{TrH}} := \mathrm{Tr}(\langle \mathbf{u}, \mathbf{v} \rangle_{\mathrm{H}})$.

  (b) For an odd $q$, $\langle \mathbf{u}, \mathbf{v} \rangle_{\mathrm{TrH}} := \mathrm{Tr}(\alpha \langle \mathbf{u}, \mathbf{v} \rangle_{\mathrm{H}})$, where $\alpha \in \mathbb{F}_q \setminus \{0\}$ is such that $\overline{\alpha} = -\alpha$.

The *Euclidean dual* (resp., *Hermitian dual* and *trace Hermitian dual*) of a code $C$ is defined to be the set

$$C^{\perp_E} := \{\mathbf{u} \in \mathbb{F}_q^n \mid \langle \mathbf{u}, \mathbf{c} \rangle_E = 0 \text{ for all } \mathbf{c} \in C\}$$

$$(\text{resp.,} \ C^{\perp_H} := \{\mathbf{u} \in \mathbb{F}_q^n \mid \langle \mathbf{u}, \mathbf{c} \rangle_H = 0 \text{ for all } \mathbf{c} \in C\}$$

$$\text{and } C^{\perp_{TrH}} := \{\mathbf{u} \in \mathbb{F}_q^n \mid \langle \mathbf{u}, \mathbf{c} \rangle_{TrH} = 0 \text{ for all } \mathbf{c} \in C\}).$$

A code $C$ of length $n$ over $\mathbb{F}_q$ is said to be *Euclidean* (resp., *Hermitian* and *trace Hermitian*) *complementary dual* if $C \cap C^{\perp_E} = \{\mathbf{0}\}$ (resp., $C \cap C^{\perp_H} = \{\mathbf{0}\}$ and $C \cap C^{\perp_{TrH}} = \{\mathbf{0}\}$).

Next proposition follows from the definitions of complementary dual codes.

**Proposition 2.1.** *Let $C$ be a code of length $n$ over $\mathbb{F}_q = \mathbb{F}_{r^2}$. Then the following statements hold.*

*i) If $C$ is a linear code, then $C$ is Euclidean complementary dual if and only if*

$$\mathbb{F}_q^n = C \oplus C^{\perp_E}.$$

*ii) If $C$ is a linear code, then $C$ is Hermitian complementary dual if and only if*

$$\mathbb{F}_q^n = C \oplus C^{\perp_H}.$$

*iii) If $C$ is an $\mathbb{F}_r$-linear code, then $C$ is trace Hermitian complementary dual if and only if*

$$\mathbb{F}_q^n = C \oplus C^{\perp_{TrH}}.$$

The following properties of codes and their duals are discussed in [4, Chapter 3].

**Proposition 2.2.** *Let $C$ be a code of length $n$ over $\mathbb{F}_q = \mathbb{F}_{r^2}$. Then the following statements hold.*

*i) If $C$ is a linear code, then $\left(C^{\perp_E}\right)^{\perp_E} = C$ and $\left(C^{\perp_H}\right)^{\perp_H} = C$.*

*ii) If $C$ is an $\mathbb{F}_r$-linear code, then $\left(C^{\perp_{TrH}}\right)^{\perp_{TrH}} = C$.*

Note that the properties $\left(C^{\perp_H}\right)^{\perp_H} = C$ and $\left(C^{\perp_E}\right)^{\perp_E} = C$ do not need to be true if $C$ is not a linear code.

The following properties are direct consequence of Proposition 2.2.

**Corollary 2.3.** *Let $C$ be a code of length $n$ over $\mathbb{F}_q = \mathbb{F}_{r^2}$. Then the following statements hold.*

*i) If $C$ is a linear code, then $n = \dim_{\mathbb{F}_q}(C) + \dim_{\mathbb{F}_q}(C^{\perp_E})$ and*

$$n = \dim_{\mathbb{F}_q}(C) + \dim_{\mathbb{F}_q}(C^{\perp_H}).$$

*ii) If $C$ is an $\mathbb{F}_r$-linear code, then $2n = \dim_{\mathbb{F}_r}(C) + \dim_{\mathbb{F}_r}(C^{\perp_{TrH}})$.*

From Corollary 2.3, to study complementary duality of codes, we focus on the Euclidean and Hermitian inner products if codes are linear, and the trace Hermitian inner product if codes are $\mathbb{F}_r$-linear over $\mathbb{F}_q$.

# 3 Characterization of Complementary Dual Subfield Linear Codes

The characterization and properties of Linear codes with Euclidean Complementary dual have been established in [3]. In this section, characterizations of Hermitian complementary dual linear codes and trace Hermitian complementary dual subfield linear codes are given in terms of orthogonal projections.

**Definition 3.1.** Let $V$ be an inner product space over a field $\mathbb{F}$. An $\mathbb{F}$-linear map $T: V \to V$ is called an $\mathbb{F}$-orthogonal projection with respect to the prescribed inner product $\langle \cdot, \cdot \rangle$ if

i) $T^2 = T$, and

ii) $\langle \mathbf{u}, \mathbf{v} \rangle = 0$ for all $\mathbf{u} \in \mathrm{Im}(T)$ and $\mathbf{v} \in \ker(T)$.

### 3.1 Characterization of Hermitian Complementary Dual Linear Codes

The following property of $\mathbb{F}_q$-orthogonal projection plays vital role in characterizing Hermitian complementary dual linear codes over $\mathbb{F}_q$

**Lemma 3.2.** *Let $C$ be a linear code of length $n$ over $\mathbb{F}_q = \mathbb{F}_{r^2}$ and let $T : \mathbb{F}_q^n \to \mathbb{F}_q^n$ be an $\mathbb{F}_q$-linear map. Then $T$ is an $\mathbb{F}_q$-orthogonal projection with respect to the Hermitian inner product onto $C$ if and only if*

$$T(\mathbf{v}) = \begin{cases} \mathbf{v} & \text{if } \mathbf{v} \in C, \\ \mathbf{0} & \text{if } \mathbf{v} \in C^{\perp_H}. \end{cases}$$

*Proof.* Suppose that $T : \mathbb{F}_q^n \to C$ is an $\mathbb{F}_q$-orthogonal projection with respect to the Hermitian inner product onto $C$. Let $\mathbf{v} \in C$ and $\mathbf{u} \in C^{\perp_H}$. Since $T$ is onto $C$, we have $C = \text{Im}(T)$. Hence, there exists a word $\mathbf{x} \in \mathbb{F}_q^n$ such that $T(\mathbf{x}) = \mathbf{v}$ and $\mathbf{v} = T(\mathbf{x}) = T^2(\mathbf{x}) = T(T(\mathbf{x})) = T(\mathbf{v})$. Since $\mathbf{u} \in C^{\perp_H}$, we have $\langle \mathbf{u}, \mathbf{v} \rangle_H = 0$ for all $\mathbf{v} \in C = \text{Im}(T)$. It follows that $\mathbf{u} \in \ker(T)$, and hence, $T(\mathbf{u}) = \mathbf{0}$.

Conversely, assume that

$$T(\mathbf{v}) = \begin{cases} \mathbf{v} & \text{if } \mathbf{v} \in C, \\ \mathbf{0} & \text{if } \mathbf{v} \in C^{\perp_H}. \end{cases}$$

Since $T$ is a function, $C \cap C^{\perp_H} = \{\mathbf{0}\}$. For each $\mathbf{v} \in \mathbb{F}_q^n$, it can be written uniquely as $\mathbf{v} = \mathbf{u} + \mathbf{w}$, where $\mathbf{v} \in C$ and $\mathbf{w} \in C^{\perp_H}$. Then $T(\mathbf{u}) = \mathbf{u}$ and $T(\mathbf{w}) = \mathbf{0}$. Hence,

$$T^2(\mathbf{u}) = T(T(\mathbf{u})) = T(\mathbf{u}) \text{ and } T^2(\mathbf{w}) = T(T(\mathbf{w})) = T(\mathbf{0}) = \mathbf{0} = T(\mathbf{w}).$$

It follows that $T^2(\mathbf{v}) = T(\mathbf{v})$ for all $\mathbf{v} \in \mathbb{F}_q^n$. Let $\mathbf{u} \in \text{Im}(T)$ and $\mathbf{v} \in \ker(T)$. Then $\mathbf{u} \in C$ and $T(\mathbf{v}) = \mathbf{0}$. It follows that $\mathbf{v} \in C^{\perp_H}$ and $\langle \mathbf{u}, \mathbf{v} \rangle_H = 0$. Hence, Im(T) and ker(T) are orthogonal with respect to the Hermitian inner product. $\square$

The characterization of Hermitian complementary dual linear codes is given as follows.

**Lemma 3.3.** Let $C$ be a linear code of length $n$ over $\mathbb{F}_q = \mathbb{F}_{r^2}$. Then $C$ is Hermitian complementary dual if and only if there exists an $\mathbb{F}_q$-orthogonal projection with respect to the Hermitian inner product from $\mathbb{F}_q^n$ onto $C$.

*Proof.* Assume that $\Pi_C$ is an $\mathbb{F}_q$-orthogonal projection with respect to the Hermitian inner product from $\mathbb{F}_q^n$ onto $C$. By Lemma 3.2, we have

$$\Pi_C(\mathbf{v}) = \begin{cases} \mathbf{v} & \text{if } \mathbf{v} \in C, \\ \mathbf{0} & \text{if } \mathbf{v} \in C^{\perp_H}. \end{cases}$$

Suppose that $C$ is not Hermitian complementary dual. Then the exists $\mathbf{u} \neq \mathbf{0}$ such that $\mathbf{u} \in C \cap C^{\perp_H}$, i.e., $\mathbf{u} \in C$ and $\mathbf{u} \in C^{\perp_H}$. It follows that $\mathbf{0} \neq \mathbf{u} = \Pi_C(\mathbf{u}) = \mathbf{0}$, a contradiction. Therefore, $C$ is Hermitian complementary dual.

Conversely, suppose $C$ is Hermitian complementary dual. Let $\mathbf{v} \in \mathbb{F}_q^n$. Then there exists a unique pair $\mathbf{u} \in C$ and $\mathbf{w} \in C^{\perp_H}$ such that $\mathbf{v} = \mathbf{u} + \mathbf{w}$. Defined a map $\Pi_C : \mathbb{F}_q^n \to \mathbb{F}_q^n$ by

$$\Pi_C(\mathbf{v}) = \mathbf{u}.$$

It is not difficult to verify that $\Pi_C$ is a linear map such that

$$\Pi_C(\mathbf{z}) = \begin{cases} \mathbf{z} & \text{if } \mathbf{z} \in C, \\ \mathbf{0} & \text{if } \mathbf{z} \in C^{\perp_H}. \end{cases}$$

Therefore, by Lemma 3.2, $\Pi_C$ an $\mathbb{F}_q$-orthogonal projection with respect to the Hermitian inner product from $\mathbb{F}_q^n$ onto $C$. $\square$

**Theorem 3.4.** Let $C$ be an $[n,k]_q$ linear code over $\mathbb{F}_q = \mathbb{F}_{r^2}$ with generator matrix $G$. Then $C$ is Hermitian complementary dual if and only if $G\overline{G}^T$ is invertible.

In this case, $\prod_C := \overline{G}^T(G\overline{G}^T)^{-1}G$ is an $\mathbb{F}_q$-orthogonal projection with respect to the Hermitian inner product from $\mathbb{F}_q^n$ onto $C$.

*Proof.* Suppose that $G\overline{G}^T$ is a non-invertible matrix. Since $G\overline{G}^T$ is a $k \times k$ matrix, we have $\text{rank}(G\overline{G}^T) < k$. It follows that

$$k = \text{null}(G\overline{G}^T) + \text{rank}(G\overline{G}^T) < \text{null}(G\overline{G}^T) + k.$$

Then $\text{null}(G\overline{G}^T) > k - k = 0$, i.e., $\{\mathbf{0}\} \neq \ker(G\overline{G}^T)$. Then there exists $\mathbf{u} \in \ker(G\overline{G}^T) \setminus \{\mathbf{0}\} \subseteq \mathbb{F}_q^k$. Hence, $\mathbf{u}G\overline{G}^T = \mathbf{0}$ and $\mathbf{u}G \in C \setminus \{\mathbf{0}\}$.

For each $\mathbf{v} \in C$, it can be written as $\mathbf{v} = \mathbf{u}'G$ for some $\mathbf{u}' \in \mathbb{F}_q^k$. Hence,

$$\langle \mathbf{u}G, \mathbf{v} \rangle_\text{H} = (\mathbf{u}G)\overline{\mathbf{v}}^T = (\mathbf{u}G)(\overline{\mathbf{u}'G})^T = \mathbf{u}G\overline{G}^T(\overline{\mathbf{u}'})^T = \mathbf{0}(\overline{\mathbf{u}'})^T = \mathbf{0}.$$

Therefore, $\mathbf{u}G \neq \mathbf{0}$ is also a word in $C^{\perp_\text{H}}$. It follows that $C \cap C^{\perp_\text{H}} \neq \{\mathbf{0}\}$, i.e., $C$ is not Hermitian complementary dual.

Conversely, assume that $G\overline{G}^T$ is invertible. Let $\mathbf{v} \in \mathbb{F}_q^n$. If $\mathbf{v} \in C$, then there exists $\mathbf{u} \in \mathbb{F}_q^k$ such that $\mathbf{v} = \mathbf{u}G$, and hence,

$$\mathbf{v}\overline{G}^T(G\overline{G}^T)^{-1}G = \mathbf{u}G\overline{G}^T(G\overline{G}^T)^{-1}G = \mathbf{u}I_k G = \mathbf{u}G = \mathbf{v}.$$

If $\mathbf{v} \in C^{\perp_\text{H}}$, then $\mathbf{v}\overline{G}^T = \mathbf{0}$, and hence,

$$\mathbf{v}\overline{G}^T(G\overline{G}^T)^{-1}G = \mathbf{0}(G\overline{G}^T)^{-1}G = \mathbf{0}.$$

Therefore, $\overline{G}^T(G\overline{G}^T)^{-1}G$ is an $\mathbb{F}_q$-orthogonal projection with respect to the Hermitian inner product from $\mathbb{F}_q^n$ onto $C$. Therefore, $C$ is Hermitian complementary dual by Lemma 3.3. □

**Example 3.5.** Let $C$ be a linear code of length $4$ over $\mathbb{F}_4 = \{0, 1, \omega, \omega^2 = \omega + 1\}$ with generator matrix $G = \begin{bmatrix} 1 & 0 & \omega & 0 \\ 0 & 1 & 1 & \omega \end{bmatrix}$. Since $G\overline{G}^T = \begin{bmatrix} 0 & \omega \\ \omega^2 & 1 \end{bmatrix}$, we have $\det(G\overline{G}^T) = 1$ which implies that $G\overline{G}^T$ is invertible. Hence, $C$ is Hermitian complementary dual by Theorem 3.4.

### 3.2 Characterization of Complementary Dual Subfield Linear Codes.

Now, we focus on the characterization of trace Hermitian complementary dual subfield linear codes.

**Lemma 3.6.** *Let $C$ be an $\mathbb{F}_r$-linear code of length $n$ over $\mathbb{F}_q = \mathbb{F}_{r^2}$ and let $T : \mathbb{F}_q^n \to \mathbb{F}_q^n$ be an $\mathbb{F}_r$-linear map. Then $T$ is an $\mathbb{F}_r$-orthogonal projection with respect to the trace Hermitian inner product onto $C$ if and only if*

$$T(\mathbf{v}) = \begin{cases} \mathbf{v} & \text{if } \mathbf{v} \in C, \\ \mathbf{0} & \text{if } \mathbf{v} \in C^{\perp \text{TrH}}. \end{cases}$$

*Proof.* Using the arguments similar to those in Lemma 3.2 and applying the trace Hermitian inner product instead of the Hermitian inner product, the statement is proved. □

**Lemma 3.7.** *Let $C$ be an $\mathbb{F}_r$-linear code of length $n$ over $\mathbb{F}_q = \mathbb{F}_{r^2}$. Then $C$ is trace Hermitian complementary dual if and only if there exists an $\mathbb{F}_r$-orthogonal projection with respect to the trace Hermitian inner product from $\mathbb{F}_q^n$ onto $C$.*

*Proof.* Assume that $\Lambda_C$ is an $\mathbb{F}_r$-orthogonal projection with respect to the trace Hermitian inner product from $\mathbb{F}_q^n$ onto $C$. By Lemma 3.6, it follows that

$$\Lambda_C(\mathbf{v}) = \begin{cases} \mathbf{v} & \text{if } \mathbf{v} \in C, \\ \mathbf{0} & \text{if } \mathbf{v} \in C^{\perp \text{TrH}}. \end{cases}$$

Suppose that $C$ is not trace Hermitian complementary dual. Then there exists $\mathbf{u} \neq \mathbf{0}$ such that $\mathbf{u} \in C \cap C^{\perp \text{TrH}}$. It follows that $\mathbf{0} \neq \mathbf{u} = \Pi_C(\mathbf{u}) = \mathbf{0}$, a contradiction. Hence, $C$ is trace Hermitian complementary dual.

Conversely, suppose that $C$ is trace Hermitian complementary dual. Let $\mathbf{v} \in \mathbb{F}_q^n$. Then there exists a unique pair $\mathbf{u} \in C$ and $\mathbf{w} \in C^{\perp \text{TrH}}$ such that $\mathbf{v} = \mathbf{u} + \mathbf{w}$. Defined a map $\Lambda_C : \mathbb{F}_q^n \to \mathbb{F}_q^n$ by

$$\Lambda_C(\mathbf{v}) = \mathbf{u}.$$

It is not difficult to see that $\Lambda_C$ is an $\mathbb{F}_r$-linear map such that

$$\Lambda_C(\mathbf{z}) = \begin{cases} \mathbf{z} & \text{if } \mathbf{z} \in C, \\ \mathbf{0} & \text{if } \mathbf{z} \in C^{\perp \text{TrH}}. \end{cases}$$

Hence, by Lemma 3.2, $\Lambda_C$ an $\mathbb{F}_r$-orthogonal projection with respect to the trace Hermitian inner product from $\mathbb{F}_q^n$ onto $C$. $\square$

**Theorem 3.8.** *Let $C$ be an $(n, r^\ell)_q$ $\mathbb{F}_r$-linear code over $\mathbb{F}_q = \mathbb{F}_{r^2}$ with generator matrix $G$. Then $C$ is trace Hermitian complementary dual if and only if $G\overline{G}^T - \overline{G}G^T$ is invertible.*

*In this case, $\Lambda_C : \mathbb{F}_q^n \to C$ defined by*

$$\Lambda_C(\mathbf{v}) = \begin{cases} \text{Tr}(\mathbf{v}\overline{G}^T)(G\overline{G}^T - \overline{G}G^T)^{-1}G & \text{if } q \text{ is even}, \\ \alpha^{-1}\text{Tr}(\alpha\mathbf{v}\overline{G}^T)(G\overline{G}^T - \overline{G}G^T)^{-1}G & \text{if } q \text{ is odd} \end{cases}$$

*is an $\mathbb{F}_r$-orthogonal projection with respect to the trace Hermitian inner product from $\mathbb{F}_q^n$ onto $C$, where $\alpha \in \mathbb{F}_q \setminus \{0\}$ is such that $\overline{\alpha} = -\alpha$.*

*Proof.* Assume that $G\overline{G}^T - \overline{G}G^T$ is not invertible. We distinguish the proof into two cases.

**Case 1** $q$ is even. Then $\text{Tr}(G\overline{G}^T) = G\overline{G}^T - \overline{G}G^T$ is not invertible. Since $\text{Tr}(G\overline{G}^T)$ is an $\ell \times \ell$ matrix, we have $\text{rank}(\text{Tr}(G\overline{G}^T)) < \ell$. It follows that

$$\ell = \text{null}(\text{Tr}(G\overline{G}^T)) + \text{rank}(\text{Tr}(G\overline{G}^T)) < \text{null}(\text{Tr}(G\overline{G}^T)) + \ell.$$

Hence, $\text{null}(\text{Tr}(G\overline{G}^T)) > \ell - \ell = 0$, i.e., $\{\mathbf{0}\} \neq \ker(\text{Tr}(G\overline{G}^T))$. Then there exists $\mathbf{u} \in \ker(\text{Tr}(G\overline{G}^T)) \setminus \{\mathbf{0}\} \subseteq \mathbb{F}_r^\ell$ such that $\mathbf{u}(\text{Tr}(G\overline{G}^T)) = \mathbf{0}$ and $\mathbf{u}G \in C \setminus \{\mathbf{0}\}$. Hence,

$$\text{Tr}(\mathbf{u}G\overline{G}^T) = (\mathbf{u}G)\overline{G}^T - \overline{\mathbf{u}G}G^T = \mathbf{u}(\text{Tr}(G\overline{G}^T)) = \mathbf{0}.$$

**Case 2** $q$ is odd. Then $\text{Tr}(\alpha G\overline{G}^T) = \alpha(G\overline{G}^T - \overline{G}G^T)$ is not invertible for all $\alpha \in \mathbb{F}_q \setminus \{0\}$ such that $\overline{\alpha} = -\alpha$. Since $\text{Tr}(\alpha G\overline{G}^T)$ is an $\ell \times \ell$ matrix, we have $\text{rank}(\text{Tr}(\alpha G\overline{G}^T)) < \ell$ and

$$\ell = \text{null}(\text{Tr}(\alpha G\overline{G}^T)) + \text{rank}(\text{Tr}(\alpha G\overline{G}^T)) < \text{null}(\text{Tr}(\alpha G\overline{G}^T)) + \ell.$$

It follows that $\text{null}(\text{Tr}(\alpha G\overline{G}^T)) > \ell - \ell = 0$, and hence, $\{\mathbf{0}\} \neq \ker(\text{Tr}(\alpha G\overline{G}^T))$. Then there exists $\mathbf{u} \in \ker(\text{Tr}(\alpha G\overline{G}^T)) \setminus \{\mathbf{0}\} \subseteq \mathbb{F}_r^\ell$ such that $\mathbf{u}(\text{Tr}(\alpha G\overline{G}^T)) = \mathbf{0}$ and $\mathbf{u}G \in C \setminus \{\mathbf{0}\}$. We have

$$\text{Tr}(\alpha \mathbf{u}G\overline{G}^T) = \alpha((\mathbf{u}G)\overline{G}^T - \overline{\mathbf{u}G}G^T) = \mathbf{u}(\text{Tr}(\alpha G\overline{G}^T)) = \mathbf{0}.$$

Altogether, it follows that $\mathbf{u}G$ is also a word in $C^{\perp \text{TrH}}$, and hence, $C \cap C^{\perp \text{TrH}} \neq \{\mathbf{0}\}$. Therefore, $C$ is not trace Hermitian complementary dual.

Conversely, assume that $\det(G\overline{G}^T - \overline{G}G^T)$. Let $\Lambda_C : \mathbb{F}_q^n \to C$ be defined by

$$\Lambda_C(\mathbf{v}) = \begin{cases} \text{Tr}(\mathbf{v}\overline{G}^T)(G\overline{G}^T - \overline{G}G^T)^{-1}G & \text{if } q \text{ is even,} \\ \alpha^{-1}\text{Tr}(\alpha \mathbf{v}\overline{G}^T)(G\overline{G}^T - \overline{G}G^T)^{-1}G & \text{if } q \text{ is odd.} \end{cases}$$

Let $\mathbf{v} \in \mathbb{F}_r^k$. If $\mathbf{v} \in C$, then there exists $\mathbf{u} \in \mathbb{F}_r^k$ such that $\mathbf{v} = \mathbf{u}G$, and hence,

$$\begin{aligned}\Lambda_C(\mathbf{v}) &= \begin{cases} \text{Tr}(\mathbf{v}\overline{G}^T)(G\overline{G}^T - \overline{G}G^T)^{-1}G & \text{if } q \text{ is even,} \\ \alpha^{-1}\text{Tr}(\alpha \mathbf{v}\overline{G}^T)(G\overline{G}^T - \overline{G}G^T)^{-1}G & \text{if } q \text{ is odd,} \end{cases} \\ &= \begin{cases} \text{Tr}(\mathbf{u}G\overline{G}^T)(G\overline{G}^T - \overline{G}G^T)^{-1}G & \text{if } q \text{ is even,} \\ \alpha^{-1}\text{Tr}(\alpha \mathbf{u}G\overline{G}^T)(G\overline{G}^T - \overline{G}G^T)^{-1}G & \text{if } q \text{ is odd,} \end{cases}\end{aligned}$$

$$\Lambda_C(\mathbf{v}) = \begin{cases} (\mathbf{u}G\overline{G}^T - \overline{\mathbf{u}G}G^T)(G\overline{G}^T - \overline{G}G^T)^{-1}G & \text{if } q \text{ is even,} \\ \alpha^{-1}\alpha(\mathbf{u}G\overline{G}^T - \overline{\mathbf{u}G}G^T)(G\overline{G}^T - \overline{G}G^T)^{-1}G & \text{if } q \text{ is odd,} \end{cases}$$
$$= \mathbf{u}(G\overline{G}^T - \overline{G}G^T)(G\overline{G}^T - \overline{G}G^T)^{-1}G$$
$$= \mathbf{u}I_k G = \mathbf{u}G = \mathbf{v}.$$

Assume that $\mathbf{v} \in C^{\perp_{\text{TrH}}}$. Then

$$\mathbf{0} = \begin{cases} \text{Tr}(\mathbf{v}\overline{G}^T) & \text{if } q \text{ is even,} \\ \text{Tr}(\alpha\mathbf{v}\overline{G}^T) & \text{if } q \text{ is odd} \end{cases}$$

and

$$\Lambda_C(\mathbf{v}) = \begin{cases} \text{Tr}(\mathbf{v}\overline{G}^T)(G\overline{G}^T - \overline{G}G^T)^{-1}G & \text{if } q \text{ is even,} \\ \alpha^{-1}\text{Tr}(\alpha\mathbf{v}\overline{G}^T)(G\overline{G}^T - \overline{G}G^T)^{-1}G & \text{if } q \text{ is odd,} \end{cases}$$
$$= \begin{cases} 0(G\overline{G}^T - \overline{G}G^T)^{-1}G & \text{if } q \text{ is even,} \\ \alpha^{-1}0(G\overline{G}^T - \overline{G}G^T)^{-1}G & \text{if } q \text{ is odd,} \end{cases}$$
$$= \mathbf{0}.$$

Hence, $\Lambda_C$ is an $\mathbb{F}_r$-orthogonal projection with respect to the trace Hermitian inner product from $\mathbb{F}_q^n$ onto $C$. Therefore, $C$ is trace Hermitian complementary dual. $\square$

**Example 3.9.** Let $\mathbb{F}_9 = \mathbb{F}_3(\omega)$, where $\omega$ is a root of $x^2 + 2x + 2$. Let $C$ be an $\mathbb{F}_3$-linear code of length 4 over $\mathbb{F}_9$ with generator matrix $G = \begin{bmatrix} 1 & 0 & \omega^3 & 0 \\ 0 & 1 & 2\omega^2 & 2 \\ \omega & 0 & \omega^4 & 0 \\ 0 & \omega & 2\omega^3 & 2\omega \end{bmatrix}$.

Since

$$G\overline{G}^T - \overline{G}G^T = \begin{bmatrix} 0 & \omega^5 & 0 & 1 \\ \omega^7 & 0 & \omega^2 & 0 \\ 0 & \omega^6 & 0 & \omega \\ 1 & 0 & \omega^3 & 0 \end{bmatrix} - \begin{bmatrix} 0 & \omega^7 & 0 & 1 \\ \omega^5 & 0 & \omega^6 & 0 \\ 0 & \omega^2 & 0 & \omega^3 \\ 1 & 0 & \omega & 0 \end{bmatrix} = \begin{bmatrix} 0 & \omega^2 & 0 & 0 \\ \omega^6 & 0 & \omega^6 & 0 \\ 0 & \omega^2 & 0 & \omega^6 \\ 0 & 0 & \omega^2 & 0 \end{bmatrix},$$

we have that $G\overline{G}^T - \overline{G}G^T$ is invertible. Hence, by Theorem 3.8, $C$ is trace Hermitian complementary dual.

**Example 3.10.** Let $\mathbb{F}_4 = \{0, 1, \omega, \omega^2 = \omega + 1\}$ and let $C$ be an $\mathbb{F}_2$-linear code of length 4 over $\mathbb{F}_4$ with generator matrix $G = \begin{bmatrix} 1 & 0 & \omega & 0 \\ 0 & 1 & 1 & \omega \\ \omega & 0 & \omega^2 & 0 \\ 0 & \omega & \omega & \omega^2 \end{bmatrix}$. Since

$$G\overline{G}^T - \overline{G}G^T = \begin{bmatrix} 0 & \omega & 0 & 1 \\ \omega^2 & 1 & \omega & \omega^2 \\ 0 & \omega^2 & 0 & \omega \\ 1 & \omega & \omega^2 & 1 \end{bmatrix} - \begin{bmatrix} 0 & \omega^2 & 0 & 1 \\ \omega & 1 & \omega^2 & \omega \\ 0 & \omega & 0 & \omega^2 \\ 1 & \omega^2 & \omega & 1 \end{bmatrix} = \begin{bmatrix} 0 & 1 & 0 & 0 \\ 1 & 0 & 1 & 1 \\ 0 & 1 & 0 & 1 \\ 0 & 1 & 1 & 0 \end{bmatrix},$$

it follow that $G\overline{G}^T - \overline{G}G^T$ is invertible. Therefore, by Theorem 3.8, C is trace Hermitian complementary dual.

# 4 Constructions of Complementary Dual Subfield Linear Codes

In this section, some constructions of complementary dual codes with respect to the Hermitian and trace Hermitian inner products are given.

### 4.1 Constructions of Hermitian Complementary Dual Linear Codes

It is well known that, for a given $[n, k, d]_q$ code, there exists an equivalent code with the same parameters such that its generator matrix is of the form $G = [I_k \ A]$ for some $k \times (n-k)$ matrix $A$ over $\mathbb{F}_q$. The generator matrix of a linear code of this form plays an important role in constructing Hermitian complementary dual codes.

The following fact is well known.

**Lemma 4.1.** *Let $p$ be a prime. Then $-1$ is a quadratic modulo $p$ if $p \equiv 1 \bmod 4$.*

Some constructions of Hermitian complementary dual subfield linear codes are given as follows.

**Theorem 4.2.** *Let $C$ be an $[n,k,d]$ linear code of length $n$ over $\mathbb{F}_q = \mathbb{F}_{r^2}$ with generator matrix $G = [I_k \ P]$. Then the following statements hold.*

*i) If $\text{char}(\mathbb{F}_q) = 2$, then a linear code $C'$ generated by matrix $G' = [I_k \ P \ P]$ is Hermitian complementary dual with parameters $[2n-k,k,d']$, where $d' \geq d$.*

*ii) If $\text{char}(\mathbb{F}_r) \equiv 1 \bmod 4$, then there exists $\lambda \in \mathbb{F}_q$ such that $\lambda^2 = -1$ and a linear code $C'$ generated by $G' = [I_k \ P \ \lambda P]$ is a $[2n-k,k,d']$ Hermitian complementary dual code, where $d' \geq d$.*

*Proof. i)* Assume that $\text{char}(\mathbb{F}_q) = 2$. Then

$$G'(\overline{G'})^T = I_k + P\overline{P}^T + P\overline{P}^T = I_k + 2P\overline{P}^T = I_k + \mathbf{0} = I_k.$$

Therefore, $G'\overline{G'}^T$ is invertible. The code $C'$ generated by $G'$ is Hermitian complementary dual by Theorem 3.4.

Since $C$ is a linear code of length $n$, $G$ has $n$ columns. Note that $P$ has $n-k$ columns. It follows that $G' = [I_k \ P \ P]$ has $k+(n-k)+(n-k) = 2n-k$ columns. Hence, $C'$ generated by $G'$ is a $[2n-k,k]$ linear code.

Next, we show that $d(C') \geq d$. Let $\mathbf{v} \in C' \setminus \{\mathbf{0}\}$. Then there exists $\mathbf{u} \in \mathbb{F}_q^k \setminus \{\mathbf{0}\}$ such that $\mathbf{v} = \mathbf{u}G' = [\mathbf{u}I_k \ \mathbf{u}P \ \mathbf{u}P]$. Hence,

$$wt(\mathbf{v}) = wt([\mathbf{u}I_k \ \mathbf{u}P \ \mathbf{u}P]) \geq wt([\mathbf{u}I_k \ \mathbf{u}P]) = wt(\mathbf{u}[I_k \ P])$$
$$= wt(\mathbf{u}G) = d(\mathbf{u}G) \geq d(C) = d.$$

Therefore, $d' = d(C') \geq d$.

*ii)* Assume that $\text{char}(\mathbb{F}_r) \equiv 1 \bmod 4$. Then $r = 4k+1$ for some integer $k$. By Lemma 4.1, there exists $\lambda \in \mathbb{F}_q$ such that $\lambda^2 = -1$. Then

$$G'(\overline{G'})^T = I_k + P\overline{P}^T + \lambda^{r+1}P\overline{P}^T$$
$$= I_k + P\overline{P}^T + \lambda^{4k+1+1}P\overline{P}^T$$
$$= I_k + P\overline{P}^T + (-1)P\overline{P}^T$$
$$= I_k.$$

Therefore, $G'\overline{G'}^T$ is invertible. Hence, by Theorem 3.4, $C'$ generated by $G'$ is Hermitian complementary dual.

Similar to $i$), we can prove that $C'$ is a $[2n-k, k, d']$ code with $d' \geq d$. □

**Example 4.3.** Let $C$ be a linear code of length 4 over $\mathbb{F}_4 = \{0, 1, \omega, \omega^2 = \omega+1\}$ with the generator matrix $G = \begin{bmatrix} 1 & 0 & \omega & 0 \\ 0 & 1 & 1 & \omega \end{bmatrix}$. Then $C$ is a $[4, 2, 2]_4$ code. By Theorem 4.2, a linear code generated by $G' = \begin{bmatrix} 1 & 0 & \omega & 0 & \omega & 0 \\ 0 & 1 & 1 & \omega & 1 & \omega \end{bmatrix}$ is Hermitian complementary dual with parameters $[6, 2, 2]_4$.

**Example 4.4.** Let $\mathbb{F}_{25} = \mathbb{F}_5(\omega)$, where $\omega$ is a root of $x^2 + 4x + 2$. Let $C$ be a linear code of length 4 over $\mathbb{F}_{25}$ with the generator matrix $G = \begin{bmatrix} 1 & 0 & \omega^{22} & \omega^5 \\ 0 & 1 & \omega^{19} & \omega^{22} \end{bmatrix}$. Then $C$ is a $[4, 2, 3]_{25}$ code. Since $2^2 \equiv -1 \bmod 5$, by Theorem 4.2, the code $C'$ generated by

$$G' = \begin{bmatrix} 1 & 0 & \omega^{22} & \omega^5 & 2\omega^{22} & 2\omega^5 \\ 0 & 1 & \omega^{19} & \omega^{22} & 2\omega^{19} & 2\omega^{22} \end{bmatrix}$$

is Hermitian complementary dual with parameters $[6, 2, d(C') \geq 3]_{25}$.

**4.2 Construction of Complementary Dual Subfield Linear Codes**

Given an $(n, r^\ell, d)_q$ $\mathbb{F}_r$-linear code $C$ over $\mathbb{F}_{q=r^2} = \mathbb{F}_r(\omega)$, a generator matrix of $C$ is an $\ell \times n$ matrix over $\mathbb{F}_q$. Using elementary row and column operations, there exists an equivalent $\mathbb{F}_r$-linear code with the same parameters with generator matrix

$$G = \begin{bmatrix} I_k & A \\ \omega I_k & \omega A \\ \mathbf{0} & B \end{bmatrix}$$

for some integer $0 \leq k \leq \dfrac{\ell}{2}$, $k \times (n-k)$ matrix $A$ over $\mathbb{F}_q$, and $(\ell - 2k) \times (n-k)$ matrix $B$ over $\mathbb{F}_q$, where $\mathbf{0}$ denotes the $(\ell - 2k) \times k$ matrix whose entries are $0$. Some constructions of trace Hermitian complementary dual codes are given as follows.

**Theorem 4.5.** *Let $C$ be an $(n, r^\ell, d)_q$ $\mathbb{F}_r$-linear code $C$ over $\mathbb{F}_{q=r^2} = \mathbb{F}_r(\omega)$ with generator matrix*

$$G = \begin{bmatrix} I_k & A \\ \omega I_k & \omega A \\ \mathbf{0} & B \end{bmatrix}$$

*such that $B\overline{B}^T - \overline{B}B^T$ is invertible. Then the following statements hold.*

*i) If $\mathrm{char}(\mathbb{F}_q) = 2$, then an $\mathbb{F}_r$-linear code $C'$ generated by*

$$G' = \begin{bmatrix} I_k & A & A & \mathbf{0} \\ \omega I_k & \omega A & \omega A & \mathbf{0} \\ \mathbf{0} & B & B & B \end{bmatrix}$$

*is trace Hermitian complementary dual with parameters $(3n - 2k, r^\ell, d')$, where $d' \geq d$.*

*ii) If $\mathrm{char}(\mathbb{F}_r) \equiv 1 \bmod 4$, then there exists $\lambda \in \mathbb{F}_q$ such that $\lambda^2 = -1$ and an $\mathbb{F}_r$-linear code $C'$ generated by*

$$G' = \begin{bmatrix} I_k & A & \lambda A & \mathbf{0} \\ \omega I_k & \omega A & \lambda \omega A & \mathbf{0} \\ \mathbf{0} & B & \lambda B & B \end{bmatrix}$$

is trace Hermitian complementary dual with parameters $(3n-2k, r^{\ell}, d')$, where $d' \geq d$.

*Proof.* In both cases, the parameters can be verified using arguments similar to those in Theorem 4.2.

Next, we show that $C'$ is trace Hermitian complementary dual.

*i)* Assume that $\text{char}(\mathbb{F}_q) = 2$. Then

$$G'\overline{G'}^T = \begin{bmatrix} I_k & \overline{\omega}I_k & \mathbf{0} \\ \omega I_k & \omega^{r+1}I_k & \mathbf{0} \\ \mathbf{0} & \mathbf{0} & B\overline{B}^T \end{bmatrix}$$

and

$$G'\overline{G'}^T - \overline{G'}G'^T = \begin{bmatrix} \mathbf{0} & (\omega+\overline{\omega})I_k & \mathbf{0} \\ (\omega+\overline{\omega})I_k & \mathbf{0} & \mathbf{0} \\ \mathbf{0} & \mathbf{0} & B\overline{B}^T - \overline{B}B^T \end{bmatrix}.$$

Since $B\overline{B}^T - \overline{B}B^T$ is invertible, $G'\overline{G'}^T - \overline{G'}G'^T$ is nonsingular. By Theorem 3.8, the $\mathbb{F}_r$-linear code $C'$ generated by $G'$ is trace Hermitian complementary dual.

*ii)* Assume that $\text{char}(\mathbb{F}_r) \equiv 1 \bmod 4$. Then $r = 4k+1$ for some $k \in \mathbb{Z}$. By Lemma 4.1, there exists $\lambda \in \mathbb{F}_q$ such that $\lambda^2 = -1$. Then $\lambda^{r+1} = \lambda^{2(2k+1)} = -1$, and hence,

$$G'\overline{G'}^T = \begin{bmatrix} I_k + (1+\lambda^{r+1})A\overline{A}^T & \overline{\omega}(I_k + (1+\lambda^{r+1})A\overline{A}^T) & (1+\lambda^{r+1})A\overline{B}^T \\ \omega(I_k + (1+\lambda^{r+1})A\overline{A}^T) & \omega^{r+1}(I_k + (1+\lambda^{r+1})A\overline{A}^T) & \omega(1+\lambda^{r+1})A\overline{B}^T \\ (1+\lambda^{r+1})B\overline{A}^T & \omega(1+\lambda^{r+1})B\overline{A}^T & (1+\lambda^{r+1})B\overline{B}^T + B\overline{B}^T \end{bmatrix}$$

$$= \begin{bmatrix} I_k & \overline{\omega}I_k & \mathbf{0} \\ \omega I_k & \omega^{r+1}I_k & \mathbf{0} \\ \mathbf{0} & \mathbf{0} & B\overline{B}^T \end{bmatrix}.$$

Consequently, we have

$$G'\overline{G'}^T - \overline{G'}G'^T = \begin{bmatrix} \mathbf{0} & (\omega+\overline{\omega})I_k & \mathbf{0} \\ (\omega+\overline{\omega})I_k & \mathbf{0} & \mathbf{0} \\ \mathbf{0} & \mathbf{0} & B\overline{B}^T - \overline{B}B^T \end{bmatrix}$$

which is invertible if and only if $B\overline{B}^T - \overline{B}B^T$ is invertible. Therefore, the $\mathbb{F}_r$-linear code $C'$ generated by $G'$ is trace Hermitian complementary dual by Theorem 3.8. □

**Example 4.6.** Let $\mathbb{F}_4 = \{0,1,\omega,\omega^2 = \omega+1\}$ and let $C$ be an $\mathbb{F}_2$-linear code of length 4 over $\mathbb{F}_4$ with the generator matrix

$$G = \begin{bmatrix} 1 & 0 & \omega & 0 \\ 0 & 1 & 0 & \omega \\ \omega & 0 & \omega^2 & 0 \\ 0 & \omega & 0 & \omega^2 \\ 0 & 0 & 1 & \omega \\ 0 & 0 & \omega^2 & 1 \end{bmatrix}.$$

Then $C$ is a $(4, 2^6, 2)_4$ $\mathbb{F}_2$-linear code.

Since $\begin{bmatrix} 1 & \omega \\ \omega^2 & 1 \end{bmatrix}\begin{bmatrix} 1 & \omega \\ \omega^2 & 1 \end{bmatrix} - \begin{bmatrix} 1 & \omega^2 \\ \omega & 1 \end{bmatrix}\begin{bmatrix} 1 & \omega^2 \\ \omega & 1 \end{bmatrix} = \begin{bmatrix} \omega+\overline{\omega} & 0 \\ 0 & \omega+\overline{\omega} \end{bmatrix}$ is invertible, the $\mathbb{F}_2$-linear code $C'$ generated by

$$G' = \begin{bmatrix} 1 & 0 & \omega & 0 & \omega & 0 & 0 & 0 \\ 0 & 1 & 0 & \omega & 0 & \omega & 0 & 0 \\ \omega & 0 & \omega^2 & 0 & \omega^2 & 0 & 0 & 0 \\ 0 & \omega & 0 & \omega^2 & 0 & \omega^2 & 0 & 0 \\ 0 & 0 & 1 & \omega & 1 & \omega & 1 & \omega \\ 0 & 0 & \omega^2 & 1 & \omega^2 & 1 & \omega^2 & 1 \end{bmatrix}$$

is Hermitian complementary dual with parameters $(8, 2^6, d(C') \geq 2)_4$ by Theorem 4.2. By direct calculation, we have $d(C') = 3$.

## Acknowledgements

This research is supported by the Thailand Research Fund under Research Grant TRG5780065.